\documentclass[english]{article}
\usepackage[T1]{fontenc}
\usepackage[latin9]{inputenc}
\usepackage[letterpaper]{geometry}
\usepackage{graphicx}

\makeatletter

\providecommand{\tabularnewline}{\\}

\newcommand{\lyxaddress}[1]{
\par {\raggedright #1
\vspace{1.4em}
\noindent\par}
}

\makeatother

\makeatother

\usepackage{babel}

\makeatother

\usepackage{babel}

\begin{document}

\title{Including Robustness in Multi-criteria Optimization for Intensity
Modulated Proton Therapy}

\author{Wei Chen, Jan Unkelbach, Alexei Trofimov, \\
 Thomas Madden, Hanne Kooy, Thomas Bortfeld, David Craft}

\maketitle

\lyxaddress{Department of Radiation Oncology, Massachusetts General Hospital
and Harvard Medical School, Boston, Massachusetts 02114, USA\\
chen.wei@mgh.harvard.edu}
\begin{abstract}
We present a method to include robustness into a multi-criteria optimization
(MCO) framework for intensity modulated proton therapy (IMPT). The
approach allows one to simultaneously explore the tradeoff between
different objectives as well as the tradeoff between robustness and
nominal plan quality. In MCO, a database of plans, each emphasizing
different treatment planning objectives, is pre-computed to approximate
the Pareto surface. An IMPT treatment plan that strikes the best balance
between the different objectives can be selected by navigating on
the Pareto surface. In our approach, robustness is integrated into
MCO by adding robustified objectives and constraints to the MCO problem.
Uncertainties (or errors) of the robust problem are modeled by pre-calculated
dose-influence matrices for a nominal scenario and a number of pre-defined
error scenarios (shifted patient positions, proton beam undershoot
and overshoot). Objectives and constraints can be defined for the
nominal scenario, thus characterizing nominal plan quality. A robustified
objective represents the worst objective function value that can be
realized for any of the error scenarios and thus provides a measure
of plan robustness. The optimization method is based on a linear projection
solver and is capable of handling large problem sizes resulting from
a fine dose grid resolution, many scenarios, and a large number of
proton pencil beams. A base of skull case is used to demonstrate the
robust optimization method. It is demonstrated that the robust optimization
method reduces the sensitivity of the treatment plan to setup and
range errors to a degree that is not achieved by a safety margin approach.
A chordoma case is analysed in more detail to demonstrate the involved
tradeoffs between target underdose and brainstem sparing as well as
robustness and nominal plan quality. The latter illustrates the advantage
of MCO in the context of robust planning. For all cases examined,
the robust optimization for each Pareto optimal plan takes less than
5 min on a standard computer, making a computationally friendly interface
possible to the planner. In conclusion, the uncertainty pertinent
to the IMPT procedure can be reduced during treatment planning by
optimizing plans that emphasize different treatment objectives, including
robustness, and then interactively seeking for a most-preferred one
from the solution Pareto surface. 

\end{abstract}

\section{Introduction}

Radiation therapy is an effective way to treat cancer by irradiating
and killing cancer cells. An ideal radiation therapy treatment delivers
sufficiently high dose to the target volume but completely spares
radiation-sensitive organs. 
Intensity modulated proton therapy (IMPT) \cite{pedroni95,lomax99}
is one of the treatment modalities that comes closest to this goal.
Compared with the exponential depth dose curve of a photon beam, the
dose deposited by a proton beam increases dramatically at the end
of range (controlled by the energy of the proton pencil beam) and
then falls off to almost zero. Compared to proton therapy delivery
based on passive scattering techniques, the comformality of the dose
distribution can be improved through IMPT. In comparison to photon
therapy the integral dose in healthy tissues sourounding the target
is greatly reduced. In IMPT, a proton beam is magnetically scanned
over the tumor volume. By using mathematical optimization techniques
to optimize the intensity of the proton beam at every location, highly
conformal treatment plans can be achieved \cite{lomax99-1}. 

However, due to the steep dose fall-off, IMPT is more sensitive to
errors than IMRT. An IMPT dose distribution can be largely distorted
even under small setup errors \cite{lomax08a}. This is inherently
due to the relatively large dose discrepancy caused by distally or
laterally shifted Bragg peaks. Another important source of uncertainty
for IMPT is range uncertainty, i.e. uncertainty of the Bragg peak
location in depth. Range uncertainty arises because the range of a
proton beam in the patient depends on the traversed tissue and is
thus flawed if the planning CT scan does not adequately represent
the patient geometry. 
See \cite{lomax08a,lomax08b} for a review on the uncertainties of
IMPT. Heuristics such as selecting the beam angles with small tissue
heterogeneity in their paths, avoiding metal implants, etc., are methods
used to increase the robustness of plans \cite{pflugfelder07}. In
photon therapy, setup uncertainty and organ motion \cite{bortfeld08,unkelbach04}
is typically accounted for using the concept of a planing target volume
(PTV). This approach is successful because a photon dose distribution
in treatment room coordinates is approximately invariant to changes
of the patient geometry. In IMPT, this approximation is not valid
in general and the usefulness of PTVs in IMPT is therefore limited.
This has recently been discussed by \cite{albertini11}.

On the other hand, robust optimization has been introduced to produce
high quality plans that are acceptable even under uncertainty. This
is achieved by incorporating the uncertainty information into the
optimization \cite{pflugfelder08,unkelbach09,fredriksson11}. Robust
optimization approaches can be probabilistic or non-probabilistic
depending on whether the underlying probability distribution of uncertainty
is known and utilized in the optimization. The work of \cite{unkelbach07,unkelbach09}
assumes a normal distribution of uncertainties, and optimizes the
expected value of a weighted sum of quadratic dose deviations. 
Without assuming a probability distribution for the uncertainty, the
works in \cite{pflugfelder08,fredriksson11} optimize treatment plans
that are as good as possible for the worst error that can occur. The
approach in \cite{fredriksson11} solves a minimax optimization problem:
it optimizes the worst score of an objective function evaluated for
a set of pre-defined error scenarios. The approach presented in \cite{pflugfelder08}
optimizes a weighted sum of two terms. The first term is the objective
function evaluated for the nominal scenario (i.e. no error occurs).
The second term is the objective function evaluated for the worst
case dose distribution, which is introduced by Lomax \cite{lomax08a}.
This worst case dose distribution is an artificial one in which every
target voxel takes the lowest dose that can occur for any error scenario,
and every healthy tissue voxel takes the highest dose. Since the worst
dose value corresponds to different error scenarios for different
voxels, the worst case dose distribution cannot be physically realized.
In contrast to the approach in \cite{pflugfelder08}, the minimax
method in \cite{fredriksson11} uses only realizable uncertainty scenarios.


Previously published works on robust optimization for IMPT share the
weakness of optimizing a weighted sum of multiple objectives, which
can lead to a tedious trial-and-error process of adjusting the weights
and redoing the expensive optimization in order to find the right
balance of objectives. MCO allows the planner to explicitly see the
tradeoff between different objectives and navigate on the Pareto surface
in real time \cite{craft05,craft11}. In \cite{chen10} we published
a fast and memory-efficient approach for optimizing IMPT in an MCO
setting. The work in this paper is an immediate extension of that
approach to robust IMPT optimization. Robust optimization is wrapped
in an MCO framework by adding robustified objectives to the MCO problem
formulation. In addition to objectives defined for the nominal scenario
as done for non-robust planning, the treatment planner can define
robustified objectives for the target and important organs at risk.
Thus, the data base of Pareto optimal treatment plans contains both
robust and non-robust treatment plans. By navigating the Pareto surface,
the treatment planner can explore the tradeoff between robustness
and nominal plan quality.

In our implementation, robust optimization performs minimax optimization
similar to the work by Fredriksson \cite{fredriksson11}. Thus, robustified
objectives correspond to worst-case objectives that return the worst
objective function value that can occur for any error scenario. By
assigning discrete probabilities to the uncertainty scenarios, our
method also allows to optimize the expected mean dose to an organ
which may be a good indicator of the protection of a parallel organ.
However, in contrast to the work by Fredriksson \cite{fredriksson11},
which uses a general non-linear constrained optimization method, we
use a customized solver for piecewise-linear convex constrained optimization.
The large-scale optimization can be solved in minutes by the projection
solver ART3+O we proposed in \cite{chen10}. The robust IMPT MCO will
be implemented in our in-house IMPT treatment planning system {}``ASTROID''
\cite{chen10} at Massachusetts General Hospital. ASTROID takes advantage
of modern multi-core computers to parallelize the multiple optimization
tasks.

The remainder of this paper is organized as follows: In section \ref{sec:methods}
we present the robust MCO framework. In section \ref{sec:skull} we
demonstrate the robust optimization method for a base-of-skull case
and compare the result to a non-robust plan optimized on a PTV. In
section \ref{sec:chordoma} we illustrate the tradeoffs involved in
robust IMPT planning for a chordoma case.

\section{Methods}

\label{sec:methods}

\subsection{Preliminaries}

We consider IMPT treatment planning for the 3D spot scanning technique
\cite{lomax99}. The spot size we model, expressed as the standard
deviation of the Gaussian dose distribution, is approximately 5 mm
at patient surface, depending on the energy layer the spot locates.
The spacing of Bragg peaks in depth is given by the proximal 80\%
to distal 80\% width of the most distal peak. This leads to a typical
spacing of the energy layers corresponding to 5-7 mm in water equivalent
range.

\subsubsection{Dose calculation}

The dose calculation of the chordoma case is done with our in-house
dose calculation algorithms for proton pencil beams at Massachusetts
General Hospital%
\footnote{The dose calculation of the base of skull case is done by an in-house
dose calculation algorithm for proton pencil beams of a finite size
at MD Anderson Cancer Center \cite{li11}.%
}. Proton pencil beam, in this context, refers both to the physical
nature of the proton beam, i.e., delivery by numerous individual narrow
proton beams, and to the computational nature of the underlying transport
model, i.e., the approximation of bulk transport as the summation
of numerous computational pencil beams. The physical model is described
in \cite{hong96}. Our implementation, however, differs. We first
transport a large set of zero-width pencil beams through the patient.
These pencil beams only model the effects of multiple Coulomb scatter.
These pencil beams, indexed by $k$, yield the dose $D_{ik}$ to each
point $i$. The physical pencil beams $j$ are computed by summing
over the mathematical pencil beams $k$ to yield the dose $D_{ij}$
from each pencil beam $j$.

\subsection{Uncertainty model}

The uncertainties in IMPT are numerous. In this paper we consider
range uncertainty and systematic setup errors. We do not consider
other potential errors like intra-fraction or inter-fraction organ
motion, which can be significant in certain sites like lung and liver.
Both range and setup errors are modeled via a discrete set of $K$
possible error scenarios. For each error scenario $I$, we use the
dose calculation engine to calculate a dose influence matrix $D_{ij}^{I}$
that stores the dose contribution of a pencil beam $j$ to voxel $i$
in error scenario $I$.

\subsubsection{Range uncertainty}

We model range uncertainty via two error scenarios: one overshoot
scenario and one undershoot scenario. Overshoot and undershoot is
modeled by a scaling of the CT Hounsfield numbers. Hence, both error
scenarios correspond to a synchronized overshoot/undershoot of all
pencil beams. This model of range uncertainty is applicable to the
components of range uncertainty that influence all proton pencil beams
in the same way. This includes, e.g., errors in the conversion of
Hounsfield numbers to relative stopping powers as well as, to some
extent, weight loss or weight gain. It is a simplification for errors
that influence different proton pencil beams in different ways (e.g.,
imaging artifacts due to metal implants).


\subsubsection{Setup uncertainty}

\label{sec:setuperror} Setup uncertainty refers to a misalignment
of the patient relative to the treatment beam. It can be modeled as
a rigid shift of the patient with respect to the isocenter. In this
paper, we characterize the magnitude of the setup error by a single
number that corresponds to the length of the three-dimensional vector
for the patient shift. It is assumed that the patient shift can occur
in any direction. For practical purpose, we represent the possible
shifts by a discrete set of equal distance shifts that are evenly
distributed on the surface of a 3D sphere. We distinguish two sets
of error scenarios: 
\begin{enumerate}
\item [K=9] Only setup errors along the three coordinate axes are considered,
i.e., anterior-posterior, left-right, and cranial-caudal. This leads
to 6 setup error scenarios. Assuming that the length of the three-dimensional
shift vector is $\lambda$ mm, those patient shifts are given by $(\pm\lambda,0,0)$
mm, $(0,\pm\lambda,0)$ mm, and $(0,0,\pm\lambda)$ mm. Together with
the nominal scenario and two range error scenarios, this yields a
total of 9 scenarios. 
\item [K=29] In addition to the above 9 scenarios, 20 additional setup
error scenarios are considered. In those 20 scenarios the patient
is shifted by $(\pm\lambda/\sqrt{2},\pm\lambda/\sqrt{2},0)$ mm, $(\pm\lambda/\sqrt{2},0,\pm\lambda/\sqrt{2})$
mm, $(0,\pm\lambda/\sqrt{2},\pm\lambda/\sqrt{2})$ mm or, $(\pm\lambda/\sqrt{3},\pm\lambda/\sqrt{3},\pm\lambda/\sqrt{3})$
mm. This yields a total of 29 scenarios. 
\end{enumerate}
We only consider setup shifts of a given length. This is motivated
by the assumption that smaller setup errors will generally lead to
smaller dosimetric errors. In the context of worst-case optimization
methods as described below, these smaller error scenarios are expected
to have none or negligible influence on the treatment plan.



\subsection{Multi-criteria robust method}

\subsubsection{General problem formulation}

In this section we formulate the multi-criteria robust optimization
problem. Let $K$ denote the number of error scenarios. The dose-influence
matrix of scenario $I$ is denoted by $D^{I}$ for $1\le I\le K$.
The nominal scenario is indexed by $I=1$. Given an intensity vector
$x$, the dose vector realized in scenario $I$ is $d^{I}=D^{I}x$,
for $1\le I\le K$. For notational convenience, let $d$ be a concatenation
of the dose vectors realized in the $K$ scenarios: $d=\left(d^{1},d^{2},\ldots,d^{K}\right)$.
Objective functions $f(d)$ and constraint functions $g(d)$ are scalar
functions of the dose vector $d$. The formulation of the robust IMPT
MCO problem is:\[
Minimize\;\left\{ f_{1}(d),f_{2}(d),\ldots,f_{M}(d)\right\} ,\]
 \[
subject\; to\; g_{i}(d)\le b_{i},\; for\; i=1,2,\ldots,N,\]
 \[
x\ge0,\]
 where $M$ is the number of objectives, $N$ is the number of constraints,
and $b_{i}$ are bounds on the values of the constraint functions.

\subsubsection{Objective and constraint functions}

We now further define the objective and constraint functions. Objective
functions can be characterized by a triple \[
\{TYPE,STRUCTURES,SCENARIOS\},\]
 where $TYPE$ refers to the functional form. In our implementation
this can be minimize the maximum dose, maximize the minimum dose,
minimize/maximize the mean dose, or minimize the ramp function (see
Subsection \ref{sub:ramp}). $STRUCTURES$ can be one structure or
a set of structures whose voxels will be involved in the function.
$SCENARIOS$ indicates whether the objective is robustified or not.
A non-robust objective function is evaluated for the nominal scenario
only and thus is a measure for nominal plan quality. A robustified
objective is evaluated for all scenarios and its value is given by
the extremum taken over the scenarios. Hence, this method implements
a worst-case optimization approach, also referred to as minimax optimization.
For example, the robustified version of the objective that minimizes
the maximum dose to an organ at risk minimizes the maximum dose to
that organ that can occur for any of the error scenarios. In addition
to nominal and robust objectives, an objective function can be evaluated
for a weighted sum of scenario doses. This is used for the mean dose
objective, which is the only objective function for which the weighted
sum of objective values for different scenarios equals the objective
function evaluated for a weighted sum of dose values. The objectives
that we use are stated explicitly in Table \ref{tab:obj}.

For the constraint functions $g$ we use exactly the same functions
we use as objectives. It should be noted that for these functions
the bounds $b_{i}$ on the right hand side of the constraints are
given in the unit Gray (Gy), and therefore have interpretable values.

\begin{table}
\begin{centering}
\begin{tabular}{c|c|c|c}
\hline 
Objective $f(d)$  & Nominal  & Robust  & Expected\tabularnewline
\hline
\hline 
Mean dose  & $\frac{1}{|V|}\sum_{i\in V}d_{i}^{1}$  & $\min/\max_{I=1}^{K}\frac{1}{|V|}\sum_{i\in V}d_{i}^{I}$  & $\frac{1}{K}\sum_{I=1}^{K}\left(w^{I}\frac{1}{|V|}\sum_{i\in V}d_{i}^{I}\right)$\tabularnewline
\hline 
Min dose  & $\min_{i\in V}d_{i}^{1}$  & $\min_{I=1}^{K}\min_{i\in V}d_{i}^{I}$  & \tabularnewline
\hline 
Max dose  & $\max_{i\in V}d_{i}^{1}$  & $\max_{I=1}^{K}\max_{i\in V}d_{i}^{I}$  & \tabularnewline
\hline 
Overdose ramp  & $\frac{1}{|V|}\sum_{i\in V}\max(0,d_{i}^{1}-d^{pres})$  & $\max_{I=1}^{K}\frac{1}{|V|}\sum_{i\in V}\max(0,d_{i}^{I}-d^{pres})$  & \tabularnewline
\hline 
Underdose ramp  & $\frac{1}{|V|}\sum_{i\in V}\max(0,d^{pres}-d_{i}^{1})$  & $\max_{I=1}^{K}\frac{1}{|V|}\sum_{i\in V}\max(0,d^{pres}-d_{i}^{I})$  & \tabularnewline
\hline
\end{tabular}
\par\end{centering}

\caption{\label{tab:obj} Objective/Constraint function definitions for a structure
that has $|V|$ voxels in volume $V$.}

\end{table}

\subsubsection{\label{sub:ramp}Ramp function }

If an OAR is located close to the target, the objectives that minimize
the maximum dose to the OAR or maximize the minimum dose to the target
are typically not sufficient to generate an acceptable treatment plan.
An objective is needed that minimizes underdose to the target even
though minimum dose is below the desired prescription dose due to
geometrical/physical reasons. We adopt the ramp function to be a powerful
complementary to the types of objectives and constraints we deal with.
An overdose (underdose) ramp function is the average overdose (underdose)
over the total number of voxels in one structure \cite{craft07}.
It is the linear analog to the standard quadratic overdose (underdose)
function. Let $d_{i},i\in V$$ $ be the dose to the voxel $i$ in
a structure $V$, $d$ be the dose vector $\left(d_{i}\right)_{i\in V}$,
and let $d^{pres}$ be the prescription dose level. The overdose ramp
function is given by: \begin{equation}
r(d,d^{pres})=\frac{1}{|V|}\sum_{i\in V}\max(0,d_{i}-d^{pres}),\label{eq:ramp}\end{equation}
 where $|V|$ is the number of voxels in the structure $V$. For any
voxel dose $d_{i}$ larger than the prescribed level $d^{pres}$,
an amount equal to the deviation is added to the overdose ramp function.
If this function is 0 it means that all voxel doses are less than
or equal to $d^{pres}$. The underdose ramp function can be similarly
defined and is useful for minimizing the underdose to a target. The
overdose ramp function applies to both target and critical structures.

\subsubsection{Combining MCO and robustness}

We introduce robustness into the MCO framework by adding robustified
objectives and constraints to the problem formulation. For example,
if minimizing the maximum dose to an organ at risk is an objective
in the conventional problem formulation, we add an additional objective
that minimizes the worst-case maximum dose to the organ at risk that
can occur for any error scenario. The advantage of this method is
that previously developed MCO methods for database generation and
Pareto surface navigation are applicable. In addition, the formulation
reflects the fact that robustness is not a global property of a treatment
plan. Instead, a treatment plan can be robust regarding the sparing
of an organ at risk but not robust regarding target coverage. In our
approach this is accounted for by adding separate robustified objectives
for different structures. In our formulation, the error scenarios
are fixed. The tradeoff between nominal plan quality and robustness
is controlled by forming convex combinations of plans optimized for
the nominal scenario and robust plans optimized for the given set
of error scenarios. Since the linear feasibility constraints stay
the same for all the optimizations, the convex combinations of any
feasible plans are still feasible.

\subsubsection{Optimization solver}

The computational challenge for this formulation is that the size
of the system matrix and therefore the number of constraints is up
to $K$ times as large as in the non-robust optimization. We solve
the problem by the fast and memory-efficient linear feasibility and
optimality solver for large scale linear programs called ART3+O \cite{chen10}.
The method is primarily designed for handling constraints that are
linear in the beam weights so that a projection onto the constraint
can be done in closed form. This is intrinsically the case for the
mean constraint and the min/max dose constraint in Table \ref{tab:obj}.
In order to project onto a ramp constraint we use an iterative heuristic
as described in the appendix. The fast convergence at sufficient precision
and nearly zero memory overhead make ART3+O among very few choices
to solve the robust IMPT MCO problem as formulated herein.

\subsubsection{MCO database generation and navigation}

After generating the anchor plans for the $M$ objectives (i.e., minimizing
each of the objectives individually), additional Pareto surface plans
can be computed as described in \cite{chen10}. Briefly, the intermediate
Pareto optimal plans can be solved by the bounded objective function
method \cite{marler04,craft06}, in which the objective values of
the average of the $M$ anchor plans are set as constraints, and another
$M$ anchor plans are generated for each of the $M$ objectives subject
to these updated constraints. The final optimal plan is a user chosen
linear combination of all the anchor plans \cite{monz08}.

\section{\label{sec:Results}Results}

We demonstrate our method on two clinical cases: a base of skull tumor
and a chordoma case. For the base of skull case a treatment plan can
be obtained without tradeoffs involved to satisfy all clinical goals.
This case is used to demonstrate the robust optimization method independent
of the MCO framework and provide a comparison with a margin based
plan. The chordoma case involves a tradeoff between brainstem sparing
and CTV coverage. We use this case to show the tradeoffs between the
target coverage, the OAR sparing and the robustness of the plan.

\subsection{\label{sec:skull}Base of skull}

\begin{figure}
\centering{}\includegraphics[scale=0.5]{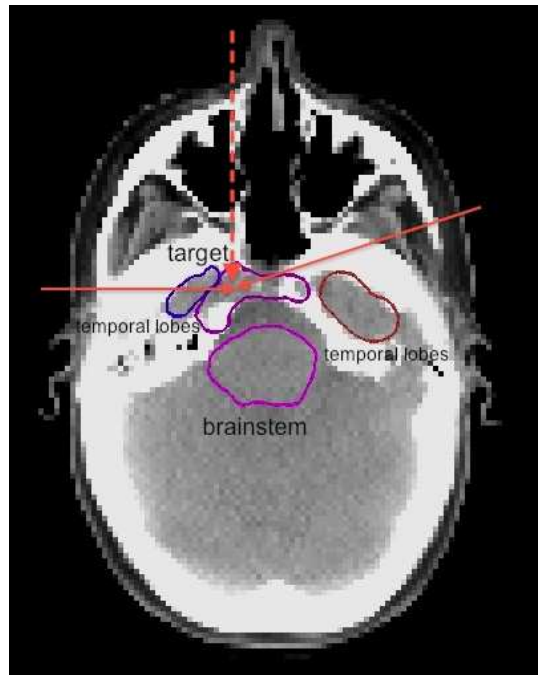}\caption{\label{fig:ct1}A slice of the CT of the base of skull case. The two
solid red arrows show the directions of the IMPT beams in the transverse
plane at gantry angles of 75 degrees and 270 degrees. The vertical
dashed red arrow shows the projection of the IMPT beam at a couch
angle of 90 degrees and gantry angle 300 degrees. The top purple structure
is the target, the bottom purple structure is the brainstem, the brown
and blue structures are the temporal lobes.}

\end{figure}

A representative CT slice of the patient is shown in Figure \ref{fig:ct1}.
Three proton beams are used in the plan for the base of skull case.
Two beams are in the transverse plane at gantry angles of 75 and 270
degrees. The third one is an obliquely incident superior-anterior
beam, which is at a couch angle of 90 degrees and gantry angle of
300 degrees. The total number of beamlets is 7,778. The dose grid
is 125$\times$125$\times$99 voxels. The voxel size is 2.5 mm $\times$
2.5 mm $\times$ 2.5 mm. We assume 3 mm setup uncertainty and 3.5\%
range uncertainty. For robust optimization we use 9 error scenarios
as defined in section \ref{sec:setuperror}. The raw dose-fluence
matrices of the 9 scenarios total to 11 Gb in size.

For this case, in order to highlight the difference between the scenario
method and the margin method, we do not use an MCO formulation. Instead,
we solve for a feasible plan that satisfies the set of constraints
given in Table \ref{tab:base of skull}. For the margin plan, the
PTV was constructed by a 3 mm isotropic expansion of the CTV. For
treatment planning, set of constraints is applied to the nominal scenario
only and the set of constraints for CTV is applied to PTV.

\begin{table}
\begin{centering}
\begin{tabular}{c|c|c|c}
\hline 
\multicolumn{4}{c}{Base of skull (Rx = 74 Gy)}\tabularnewline
\hline 
Structures  & Type  & Scenarios  & Bound (in Gy)\tabularnewline
\hline
\hline 
CTV  & min  & all  & $\ge70$\tabularnewline
\hline 
CTV  & max  & all  & $\le80$\tabularnewline
\hline 
CTV  & min  & nominal  & $\ge74$\tabularnewline
\hline 
CTV  & max  & nominal  & $\le77$\tabularnewline
\hline 
CTV  & overdose ramp ($d^{pres}$ = Rx)  & nominal  & $\le0.5$\tabularnewline
\hline 
brainstem  & max  & all  & $\le60$\tabularnewline
\hline 
brainstem  & mean  & nominal  & $\le10$\tabularnewline
\hline 
brainstem  & mean  & weighted (see caption)  & $\le11$\tabularnewline
\hline 
optic chiasm  & max  & all  & $\le62$\tabularnewline
\hline 
brain, brain-CTV, R/L temporal lobe  & mean  & nominal  & $\le30$\tabularnewline
\hline 
all structures  & max  & all  & $\le90$\tabularnewline
\hline
\end{tabular}
\par\end{centering}

\caption{\label{tab:base of skull}Constraints of the robust method for the
base of skull case. The {}``weighted'' scenario is the nominal scenario
with weight 0.2 and the other 8 scenarios with weight 0.1. }

\end{table}

\begin{figure}
\begin{centering}
\includegraphics[scale=0.5]{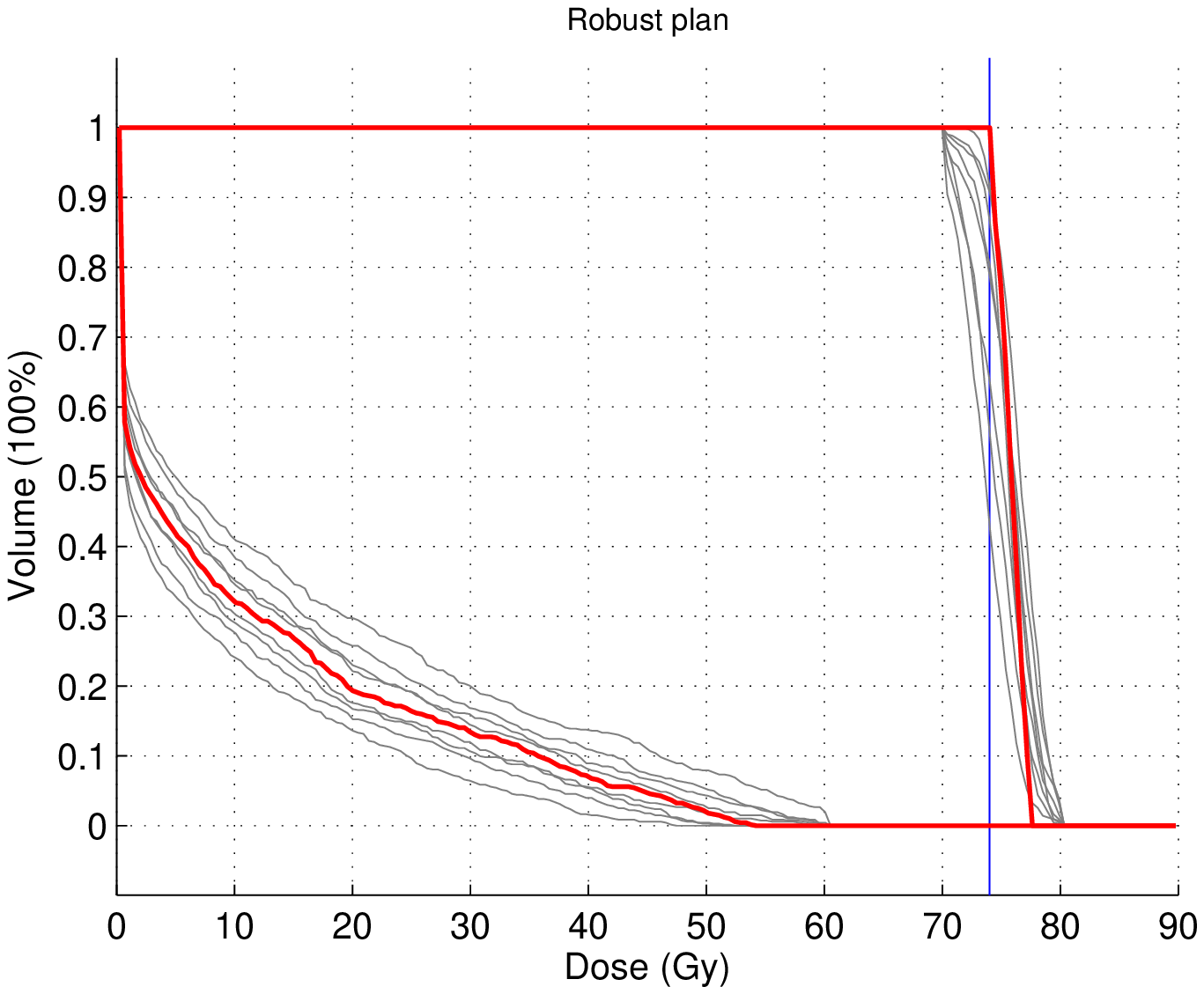}\includegraphics[scale=0.5]{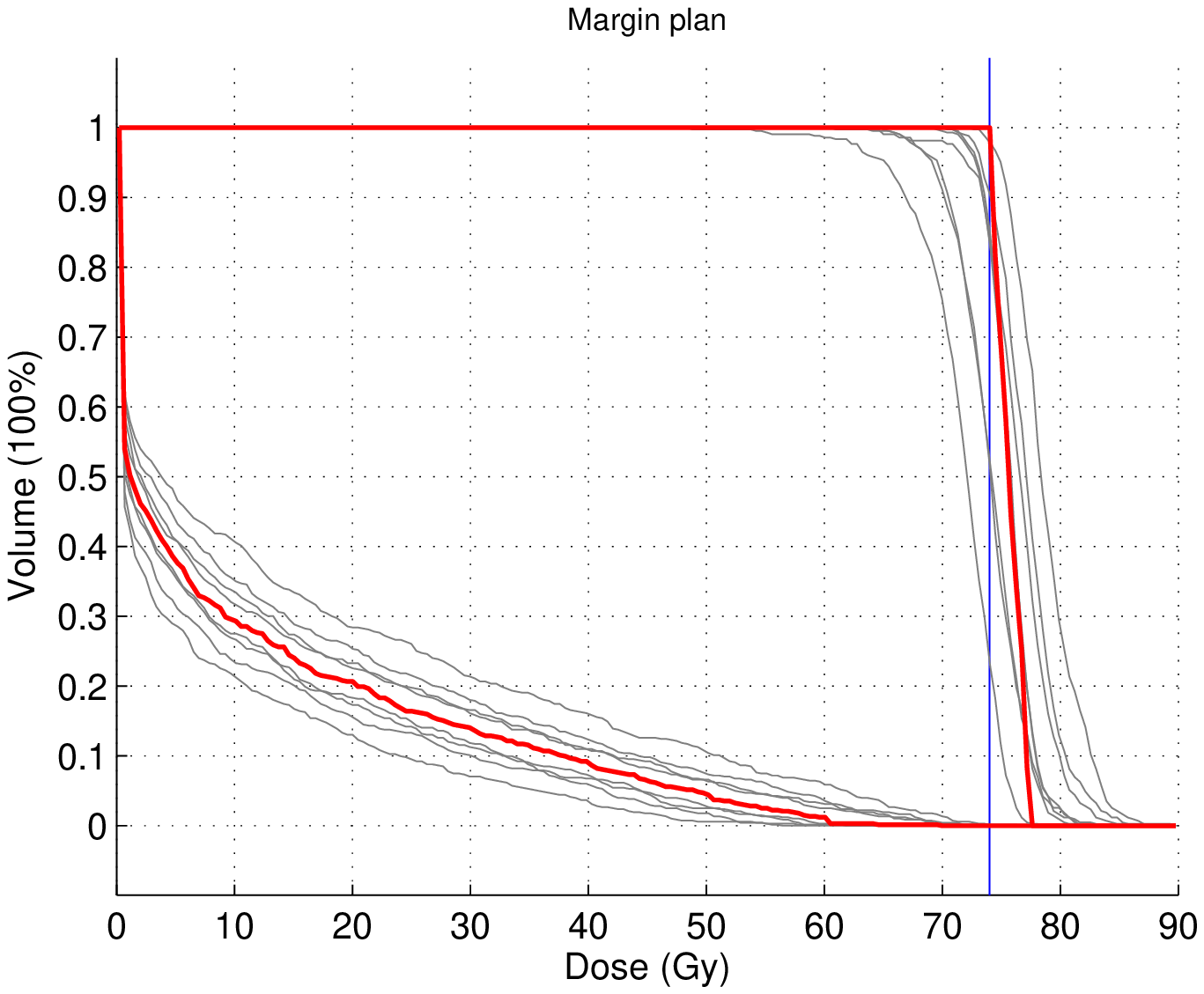} 
\par\end{centering}

\caption{\label{fig:skull-robust-ptv}The DVHs of CTV and brainstem for the
9 scenarios. The DVH for the nominal scenario is in red and the DVHs
for 8 error scenarios are in gray. Left is the robust plan, right
is the margin plan. The vertical blue line is the prescribed dose
to the target.}

\end{figure}

Figure \ref{fig:skull-robust-ptv} compares the DVHs of the CTV and
the brainstem of the robust plan and the margin plan evaluated in
all 9 scenarios. For both plans, the nominal DVHs (the red lines)
are almost equally good. They both satisfy all constraints. However,
the target coverage of the margin plan is much worse than the robust
plan due to large underdose and overdose in some of the error scenarios.
And unlike the margin plan, the robust plan restricts the maximum
brainstem dose to 60 Gy in all the scenarios. This comparison demonstrates
that by incorporating range and setup uncertainty information in the
optimization, the sensitivity of IMPT plans against errors can be
reduced.



%

\subsection{\label{sec:chordoma}Chordoma}

\subsubsection{Patient geometry, problem size and Calculation time}

\begin{figure}
\centering{}\includegraphics[scale=0.5]{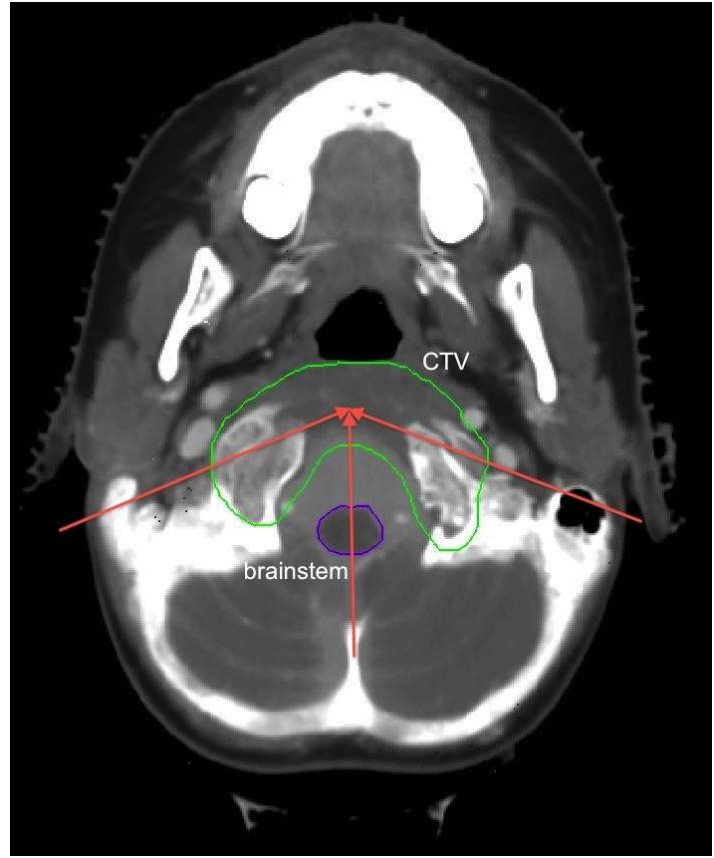}\caption{\label{fig:ct2}A slice of the CT of the chordoma case. The three
red arrows show the directions of the three IMPT beams in the transverse
plane at gantry angles of 110 degrees, 180 degrees and 250 degrees.
The green structure is the CTV and the purple structure is the brainstem.}

\end{figure}

Three proton beams are used in the plan for this chordoma case. They
are in the transverse plane at gantry angles of 110 degrees, 180 degrees
and 250 degrees (see the CT image in Figure \ref{fig:ct2}). The total
number of beamlets is 9,623. The dose grid is 88$\times$103$\times$77
voxels. The voxel size of the dose grid is 2 mm $\times$ 2 mm $\times$
2.5 mm. We assume 3 mm setup uncertainty and 5\% range uncertainty.
The raw dose-fluence matrices of the 29 scenarios used for this case
total to 16.3 Gb in size. Each individual Pareto optimal plan optimization
takes up to 5 minutes on a single 2.66G Intel Xeon CPU.

\begin{table}
\begin{centering}
\begin{tabular}{c|c|c|c}
\hline 
\multicolumn{4}{c}{Chordoma (Rx = 78 Gy)}\tabularnewline
\hline 
Structures  & Type  & Scenarios  & Bound (in Gy) / Direction\tabularnewline
\hline
\hline 
\multicolumn{4}{c}{Objectives}\tabularnewline
\hline 
CTV  & underdose ramp ($d^{pres}$ = Rx)  & all  & minimize\tabularnewline
\hline 
brainstem, spinal cord  & max  & all  & minimize\tabularnewline
\hline 
\multicolumn{4}{c}{Constraints}\tabularnewline
\hline 
CTV  & min  & nominal  & $\ge60$\tabularnewline
\hline 
CTV  & max  & nominal  & $\le85.8$\tabularnewline
\hline 
R/L cochlea  & max  & 9  & $\le50$\tabularnewline
\hline 
R/L parotid  & mean  & 9  & $\le26$\tabularnewline
\hline
\end{tabular}
\par\end{centering}

\caption{\label{tab:chordoma}Objectives and constraints of the robust method
for the chordoma case. The brainstem and cord are unioned for the
objective.}

\end{table}

\subsubsection{Analysis of tradeoffs}

For this case, the two main conflicting objectives are tumor coverage
and brainstem sparing. Those objectives are implemented via a maximum
dose objective for the brainstem and an underdose ramp objective for
the CTV. All other critical structures are handled via constraints.
All dose constraints and objectives we use are summarized in Table
\ref{tab:chordoma}.

\begin{figure}
\centering{}\includegraphics[scale=0.75]{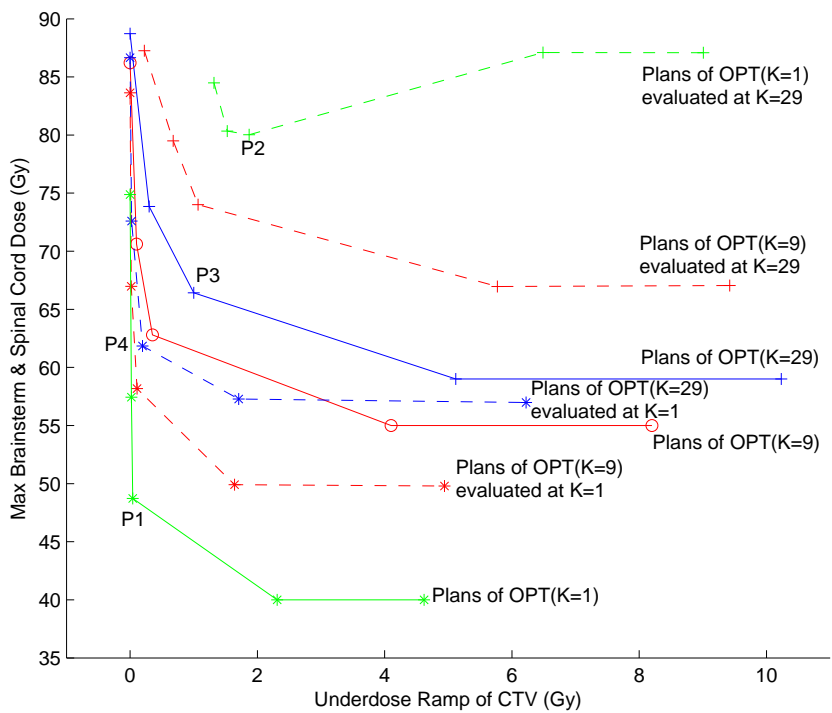}\caption{\label{fig:pareto}Pareto surfaces of the chordoma case for three
scenario set sizes $K=1,9,29$, and the same plans evaluted for different
scenario sets. The solid green line is the Pareto surface when the
two objectives only apply to the nominal scenario ($K=1$). The solid
red line is the Pareto surface when the two objectives are robustified
by the 9 scenarios ($K=9$). The solid blue line is the Pareto surface
when the two objectives are robustified by the 29 scenarios ($K=29$).
The dashed green line shows the 5 plans on the $K=1$ Pareto surface
evaluated at 29 scenarios. The dashed blue line is the 5 plans on
the $K=29$ Pareto surface evaluated at the nominal scenario. The
upper dashed red line is the 5 plans on the $K=9$ Pareto surface
evaluated at 29 scenarios. The lower dashed red line is the 5 plans
on the $K=9$ Pareto surface evaluated at the nominal scenario.}

\end{figure}

\begin{figure}
\begin{centering}
\includegraphics[scale=0.5]{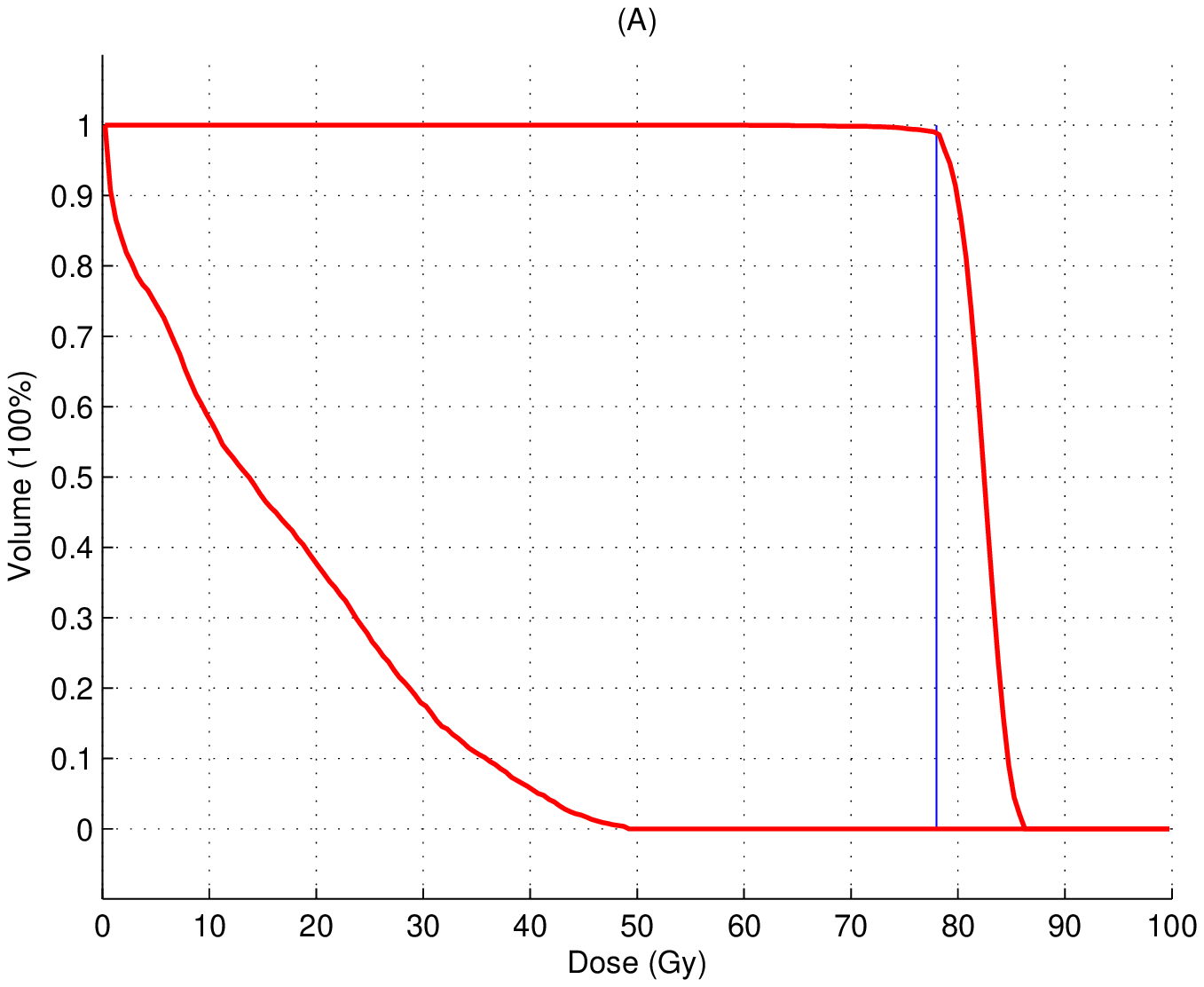}\includegraphics[scale=0.5]{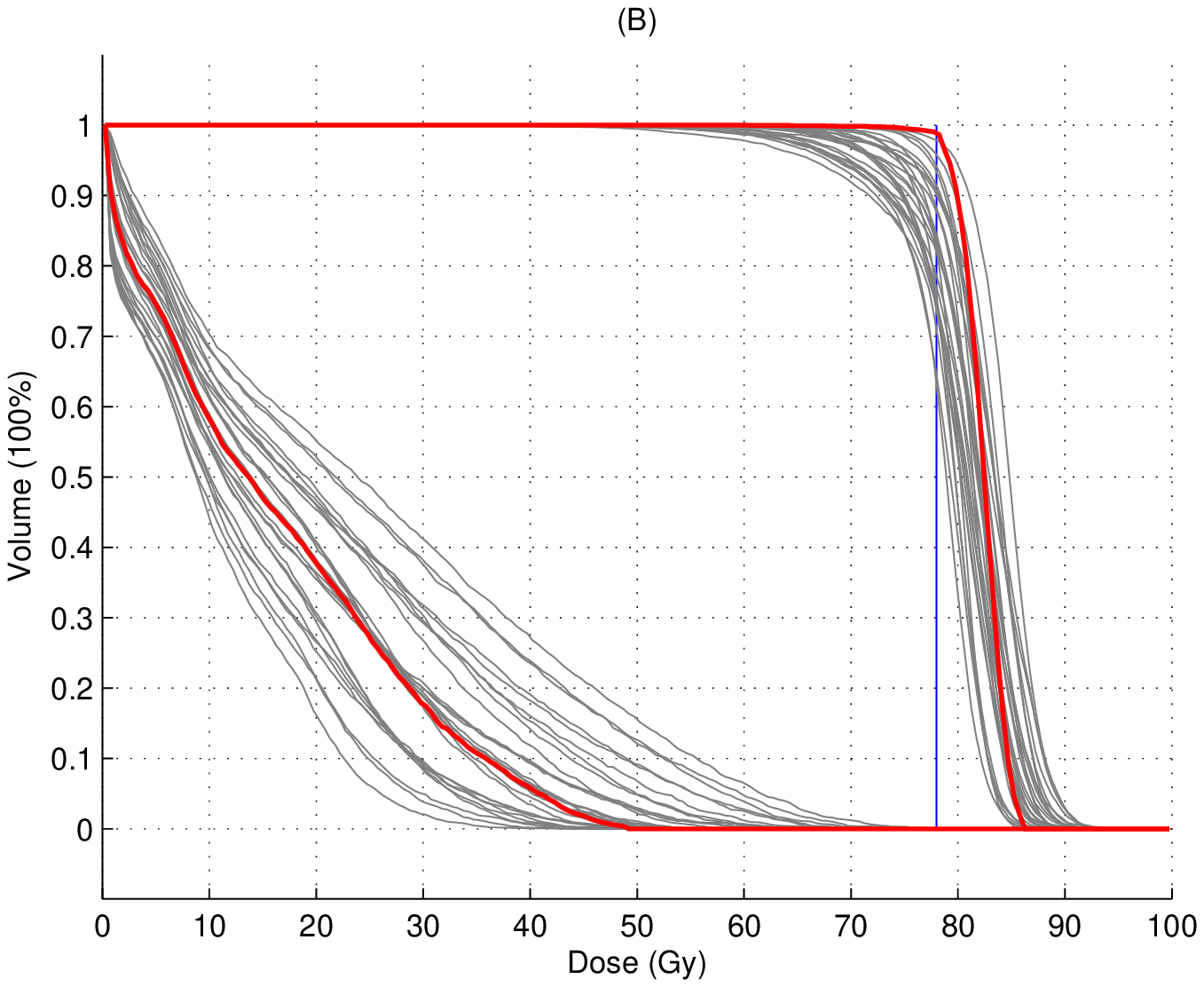}\\
 \includegraphics[scale=0.5]{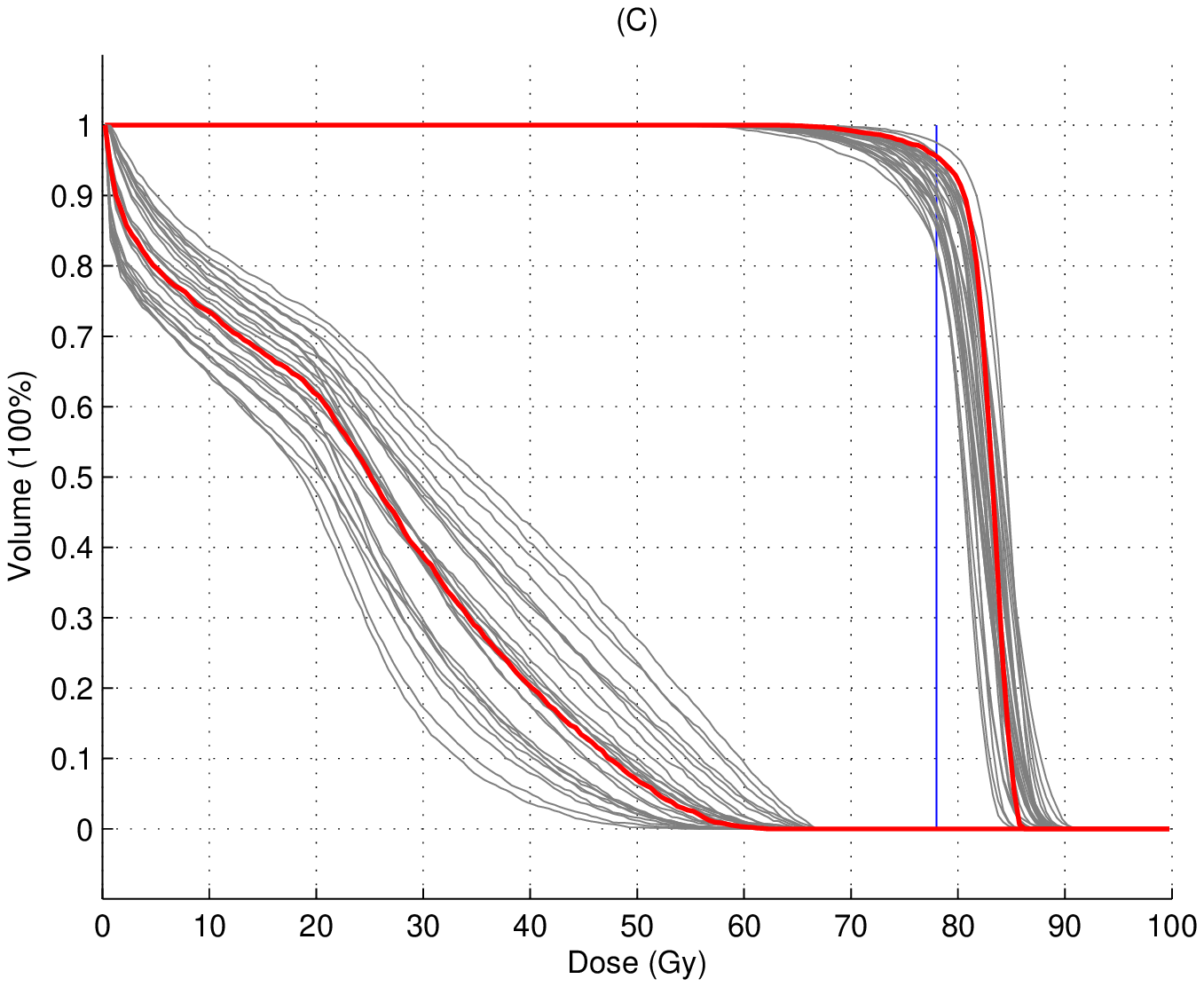}\includegraphics[scale=0.45]{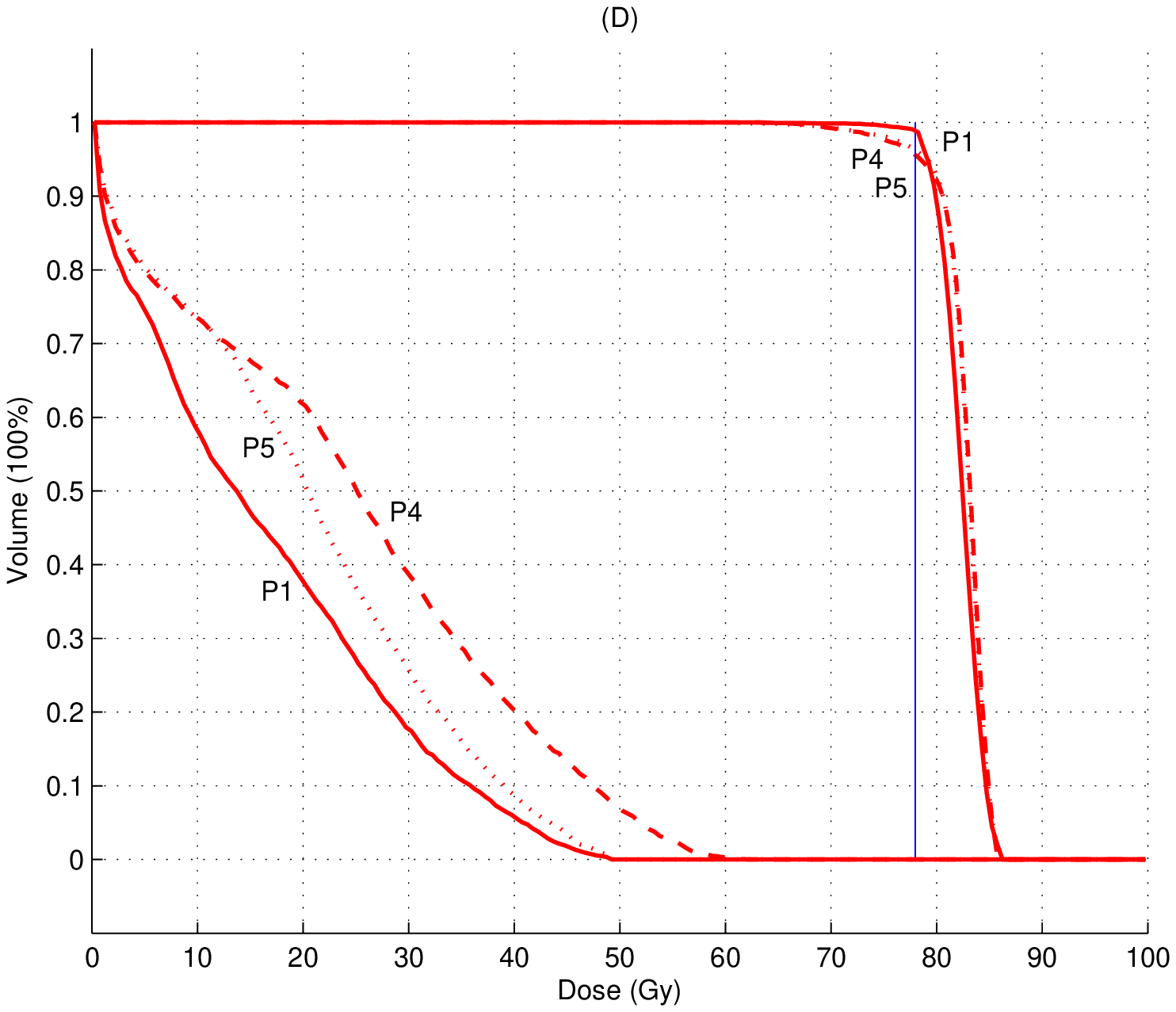} 
\par\end{centering}

\caption{\label{fig:story}(A) The nominal DVH of the non-robust plan (P1 in
Figure \ref{fig:pareto}). (B) The DVHs in 29 scenarios of the non-robust
plan (P2 in Figure \ref{fig:pareto}). (C) The DVHs in 29 scenarios
of the 29 scenarios optimized robust plan (P3 in Figure \ref{fig:pareto}).
(D) Comparison of the nominal DVHs for three plans: solid line: nominal
plan (P1 in Figure \ref{fig:pareto}), dashed line: robust plan (P4
in Figure \ref{fig:pareto}), dotted line: robust plan with optimized
nominal brainstem dose (P5 in Figure \ref{fig:pareto2}). Nominal
scenario DVHs are in red and error scenarios are in gray.}

\end{figure}

We start the discussion of this case by characterizing the tradeoff
between target coverage and brainstem sparing for the nominal, non-robust
case. The solid green line in Figure \ref{fig:pareto} shows the Pareto
surface that corresponds to the optimization problem in Table \ref{tab:chordoma}
(except that objectives and constraints are evaluated for the nominal
scenario only). Here, the Pareto surface has been approximated by
the two anchor plans plus three intermediate plans. The treatment
plan labeled as P1 represents a good tradeoff between the two objectives.
The corresponding DVHs for CTV and brainstem for the nominal scenario
are shown in Figure \ref{fig:story}(A), indicating that satisfying
both CTV coverage and brainstem sparing can be achieved.

If we evaluate this treatment plan for the 29 error scenarios defined
in Subsection \ref{sec:setuperror}, we observe that the treatment
plan quality is insufficient for several error scenarios. This is
demonstrated in Figure \ref{fig:story}(B), which shows the DVHs for
CTV and brainstem for the 29 error scenarios. In order to visualize
the degradation of plan quality in Figure \ref{fig:pareto}, we can
evaluate the robustified objectives (i.e., the maximum brainstem dose
that occurs in any of the 29 scenarios and the worst-case underdose
ramp value for the CTV) for the 5 treatment plans that approximate
the Pareto surface for the nominal optimization problem. The result
is given by the dashed green line in Figure \ref{fig:pareto}.

In order to obtain a treatment plan that is robust against errors,
we can calculate the Pareto surface for the robustified objectives,
i.e., the only two objectives are given by the robustified brainstem
maximum dose and the robustified CTV underdose ramp. Nominal plan
quality is not explicitly incorporated into treatment plan optimization.
The result is shown as the solid blue line in Figure \ref{fig:pareto}.
Comparing the solid blue line to the dashed green line indicates that
the robustness of the treatment plan could be improved through robust
optimization. This is also illustrated via the DVHs for the treatment
plan labeled as P3. Figure \ref{fig:story}(C) shows DVHs for the
29 error scenarios; a comparison to Figure \ref{fig:story}(B) reveals
the improved robustness.

Another important interpretation of the blue and the green Pareto
surface (solid lines) in Figure \ref{fig:pareto} is that the tradeoff
between brainstem sparing and CTV coverage becomes harder if we include
robustness. In the nominal case (solid green line) it is unproblematic
to determine a treatment plan that fulfills clinical goals. As soon
as robustness is enforced, the tradeoff is harder, meaning that, e.g.,
worse target coverage has to be accepted in order to maintain a given
level of brainstem sparing.

We further observe that the improvement in plan robustness is associated
with a deterioration of nominal plan quality. This can be visualized
if we evaluate the nominal objectives for the 5 robust treatment plans
that define the solid blue line in Figure \ref{fig:pareto}. The result
is given by the dotted blue line. The compromised nominal plan quality
is also illustrated in Figure \ref{fig:story}(D), which shows the
DVHs for the nominal scenario for the treatment plan labeled as P4. 

In the following paragraphs we further analyse the tradeoff between
robustness and nominal plan quality. Intuitively, we can assume that
there is a tradeoff between the nominal brainstem dose and robustness
of CTV coverage. If we want to cover the CTV under uncertainty, it
is likely that we have to accept higher brainstem doses even in the
nominal case. Looking at Figure \ref{fig:pareto} we can ask the following
question: To what extent can the nominal brainstem dose be improved
by slightly worsening the robustness in CTV coverage?

Generally, we can consider the treatment planning problem as an MCO
problem with four objectives: (1) nominal brainstem maximum dose,
(2) robustified brainstem maximum dose, (3) nominal CTV underdose
ramp, and (4) robustified CTV underdose ramp. The treatment plans
on the solid blue and green lines in Figure \ref{fig:pareto} are
all Pareto optimal in this four-objective problem. However, they are
extreme plans in the sense that they consider either the nominal objectives
or the robustified objectives, but not both simultaneously.

\begin{figure}
\centering{}\includegraphics[scale=0.5]{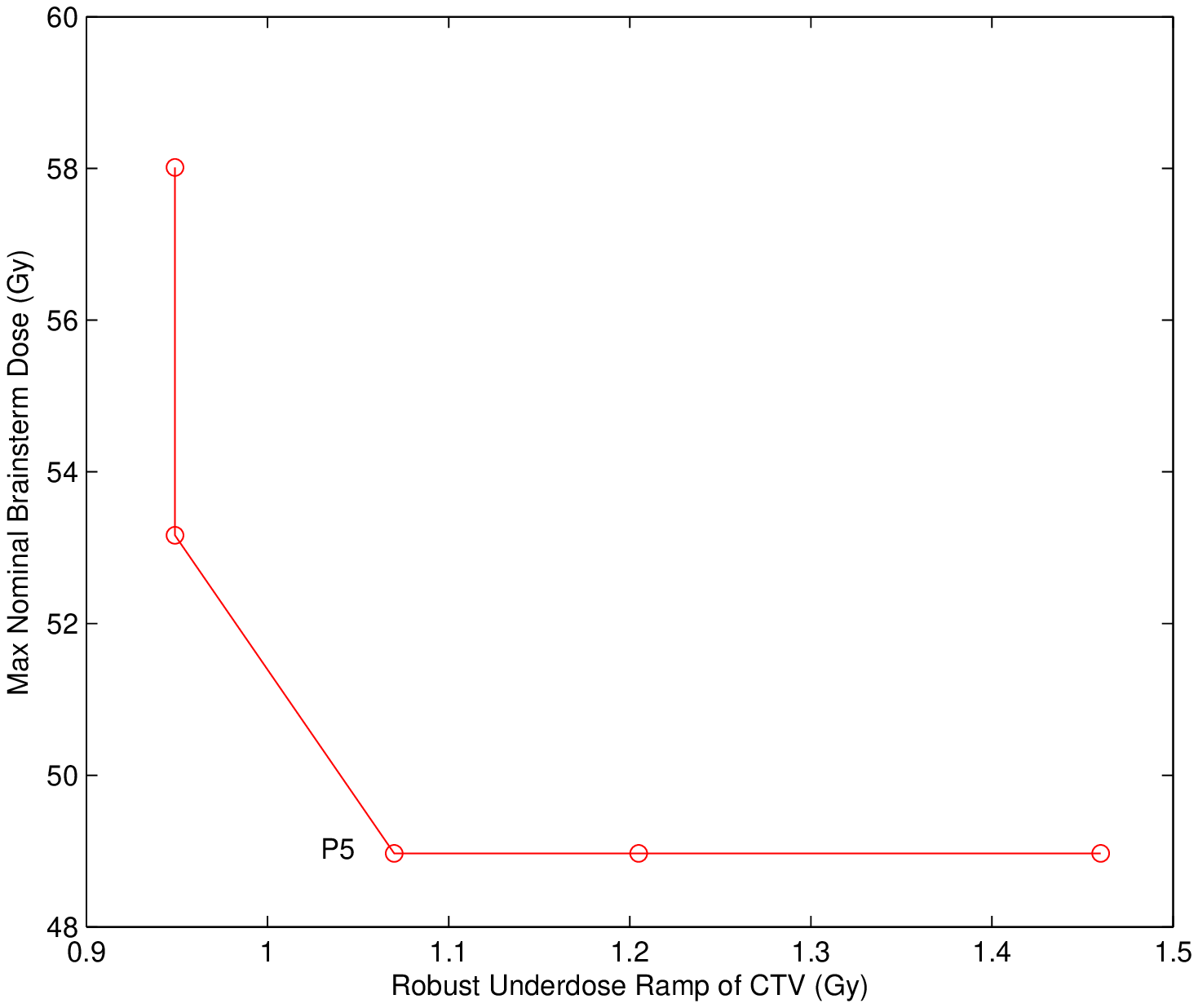}\caption{\label{fig:pareto2}The Pareto surface visualizing the trade-off between
nominal brainstem dose and robust underdose ramp of the CTV.}

\end{figure}

We now want to visualize the tradeoff between nominal brainstem dose
and robustness of CTV coverage. In order to do this, we start off
with the treatment plan labeled P3. The location of P3 in the diagram
of Figure \ref{fig:pareto} yields optimal objective values for the
robustified objectives. The point P4, which corresponds to the same
treatment plan, indicates the corresponding values of the nominal
objectives. We now impose a constraint on the value of the robustified
brainstem maximum dose objective and the nominal CTV underdose ramp
objective. The constraint level is given by P3 and P4, i.e., approximately
67 Gy and 0.2 Gy, respectively. Given these constraints, we can calculate
the two-dimensional Pareto surface that characterizes the tradeoff
between nominal brainstem dose and robust CTV coverage. The result
is shown in Figure \ref{fig:pareto2}.

It is apparent that one can reduce the nominal brainstem maximum dose
substantially without worsening the robustness of the CTV coverage.
Whereas the nominal brainstem dose is 62 Gy for the plan P4 in Figure
\ref{fig:pareto}, it can be reduced to at least 53 Gy without worsening
the robust CTV objective noticeably. The nominal DVHs for the plan
labeled P5 in Figure \ref{fig:pareto2} is shown as the dotted line
in Figure \ref{fig:story}(D).

\subsubsection{The number of error scenarios}

Above, we discussed the Pareto surface for robust treatment plans
that were optimized for 29 error scenarios. Figure \ref{fig:pareto}
also shows the Pareto surface of robust plans that are optimized using
the smaller set of $K=9$ error scenarios (red solid line). In addition,
the dashed red line shows 5 plans on the red Pareto surface when the
values of the robustified objective functions are evaluated for 29
error scenarios. Similarly, the dotted red curve shows the reevaluation
of the same plans for the nominal scenario. The discrepancy between
the solid line and the dashed line indicates that if a plan is robustified
only against 9 error scenarios, there may be error scenarios among
the larger set of 29 scenarios that show larger maximum brainstem
doses or worse target coverage. On the other hand, not robustifying
against the additional 20 setup error scenarios leads to better nominal
plan quality (dotted line).

\section{Discussion and conclusion}

Compared to photon radiation therapy, proton therapy offers the chance
for superior dose distributions due to the shape of the dose deposition
curve of a proton pencil beam. The exact location of the dose peak
depends on the incident pencil beam energy and the (line integral
of) stopping powers along the proton pencil beam path. In IMPT, the
intensity of thousands of pencil beams, incident from a small number
of angles, are optimized to collectively produce a dose distribution
that best conforms to the target and spares critical structures. Due
to the sensitivity of the individual Bragg peaks to the amount and
type of physical matter that each pencil beam passes through, an optimized
dose distribution can be highly sensitive to errors in patient setup
and discrepancies in the actual versus predicted range of the Bragg
peaks. Robust optimization provides a numerical technique to compute
plans (i.e., the fluence contribution from each pencil beam) that
are less sensitive to these possible errors.

It is not trivial to determine what to strive for regarding a {}``robust\textquotedblright{}
plan. For one, a plan that robustly covers the target will necessarily
produce more dose to critical structures. Conversely, a plan that
robustly spares a nearby critical structure will likely underdose
the target. Additionally, the level of robustness needs to be determined,
and this may be best decided only after observing how robustness changes
the overall plan quality. Such observations indicate that an interactive
system, where a user can explore various robustness options and levels,
will be useful. In the present work we have presented such a system
in the form of Pareto-surface based MCO utilizing multiple scenarios
to model IMPT delivery errors.

\subsection{Robust optimization: exploiting the redundancy of constraints}

Compared to single scenario optimization, multiple scenario optimization
has a greatly increased problem size and therefore will in general
stress the computational environment of this already large problem
(millions of voxels and thousands of pencil beams). In terms of the
computational burden of requiring a large number of error scenarios
$K$, the projection solver that we use, ART3+O, is well-suited. Computational
burden (used loosely to mean both memory requirements and computation
time) of typical gradient-based and linear programming algorithms
scale at least linearly with the input problem size. In the case of
ART3+O, only active constraints are in the working memory, and while
more scenarios add more constraints to the problem, only a fraction
of those constraints are binding for a given optimization run. For
example, a worst-case (i.e. all scenarios) constraint on a spinal
cord maximum dose level will be dominated by the scenario(s) that
have the spinal cord shifted in the direction of the target and/or
the Bragg peaks shifted into the spinal cord. Thus even though the
constraints will be written to constrain the spinal cord doses for
all scenarios, most of these constraints will be redundant and only
the constraints for selected scenarios will be important. The ART3+O
solver naturally exploits such redundancy of constraints, as described
in more detail in \cite{chen10}.

\subsection{Combining robust optimization with MCO}

Given a solver capable of handling large IMPT instances and the observation
that the notion of robustness in radiotherapy planning is inherently
a question of trading off between nominal plan quality and robustified
plan quality, we have implemented and demonstrated a robust MCO system
that pre-computes a set of Pareto optimal plans as a way to expose
these tradeoffs to treatment planners. To our knowledge, this is the
first such system to be described.

In Subsection \ref{sec:chordoma} for the chordoma case, we have gone
through a typical tradeoff analysis that a physicist might experience
in the course of navigating the Pareto surface. Although we have not
described a navigation system in this paper, the technique detailed
in \cite{monz08} is applicable to this setting. The concept of using
weighted averages of pre-computed Pareto optimal plans to approximate
the continuous Pareto surface is well fitted to IMPT planning, where
the average of multiple plans is deliverable (i.e., the fluence map
sequencing step that is needed in IMRT planning, and adds a complication
to the navigation procedure, is not needed here).

In our approach to robust MCO, the error scenarios are fixed. The
tradeoff between nominal plan quality and robustness is controlled
by forming convex combinations of plans optimized for the nominal
scenario and robust plans optimized for the given set of error scenarios.
An alternative approach to controlling plan robustness would be to
vary the magnitude of the error. This approach is however not pursued
here, the reason being that existing methods for database generation
and navigation are not applicable to such an approach.

A limitation in the current implementation of the projection solver
is the Pareto optimal plan database generation method. In its current
form, the solver is built to optimize single objectives, which is
why we generate interior Pareto surface plans by adjusting constraint
levels for a set of the objectives and then minimizing another one
of them. More advanced Pareto surface generation strategies \cite{rennen11,bokrantz11},
which compute additional interior points on the surface in such a
way to minimize the gap between the upper and lower Pareto surface
bounds, require solvers to be able to optimize weighted sums of the
underlying objectives. Using the solver ART3+O for weighted sum minimization
requires the introduction of auxiliary variables (the details of such
techniques are well known in the linear optimization literature, see
for example \cite{bertsimas97}). While this is straightforward, it
has not been pursued at this point.

\subsection{Remarks}

The hard constraints set up for the constrained optimization problem
define the feasible plan space where the planner will later navigate.
Therefore those constraint bounds should be selected loosely to make
room for the navigation to find a satisfactory Pareto optimal plan.
Of course, more user experience on the specific disease site will
help in determining typical values of those constraint bounds.

For simplicity of presentation, we have ignored the clinical IMPT
concepts of single field uniform dose (SFUD) and fraction groups.
SFUD is an additional constraint that specifies that the dose to a
target from a single beam needs to be uniform, within some specified
tolerance. This greatly restricts the degrees of freedom in IMPT planning;
it is used clinically to produce more robust plans without explicitly
performing robust optimization. Using auxiliary variables it is straightforward
to include such a constraint in the system, and indeed our initial
release of ASTROID includes this option. 

In this work we have used both $K=9$ and $K=29$ scenarios, and we
observe that the two choices result in different Pareto surfaces.
Since one dose-fluence matrix is computed for each scenario, it is
desirable computationally to keep $K$ as low as possible, but in
general the choice of $K$ will depend on the disease site and the
uncertainty settings (magnitude of setup and range errors modeled).
Further experimentation is required to be able to make statements
about how many scenarios are needed. Furthermore, investigating the
number of error scenarios required for optimization is linked to the
question how the robustness of a treatment plan should be measured
and visualized. Minimax optimization by its nature aims at optimizing
treatment plans for worst case that is considered in the uncertainty
model. In practice, however, a focus on the worst case may not always
be desired. Instead, the clinical goal may be that a treatment plan
is acceptable for the majority of patients, suggesting statistical
measures for robustness evaluation. It can be hypothesized that a
large number of error scenarios is needed if the worst case scenario
is criterion for robustness. Instead, a smaller number of scenarios
may be sufficient if average plan quality is the criterion of choice.

The questions regarding the best robustness evaluation measure, the
number of error scenarios used in robust optimization, the magnitude
of the assumed errors, and the type of robust optimization method
that is to be used are all linked. Gaining further experience in clinical
IMPT planning may be needed to address these questions. Partly for
this reason, from a system design point of view, we have opted for
{}``user has complete control\textquotedblright{}. Because these
are the early days of IMPT, and it is not clear exactly how much emphasis
should be put on robust plan quality versus nominal plan quality,
we opt for a system that allows planners to view the spectrum of possibilities.
The point of this paper is to introduce a system that is capable of
studying these issues, not to answer them. It is likely that they
can only be answered in a site specific (or even patient specific)
way, and that the answers depend on the characteristics of the pencil
beam scanning system under consideration.

\subsection{Summary}

In summary, we presented a new method for robust IMPT optimization.
In our approach, uncertainty is modeled by a discrete set of error
scenarios. For each error scenario, a separate dose-influence matrix
is pre-computed to calculate the dose distribution under those errors.
The current implementation performs minimax optimization, i.e. treatment
plans are optimized such that an objective is minimized for the worst
error scenario that can occur. Our solver is customized for linear
constraints and exploits the redundancy in the constraint set that
is inherent to minimax optimization. In addition, we present an approach
to incorporate robustness into a multi-criteria optimization. The
approach can take advantage of existing methods for database generation
and Pareto surface navigation. In the context of robust optimization,
there is a special need for MCO planning methods, first because tradeoffs
between different volumes of interests become harder, and second because
it leads to a tradeoff between robustness and nominal plan quality.
This has been illustrated for a Chordoma case.

\section*{ACKNOWLEDGMENTS}

The authors thank Wei Liu and Xiaodong Zhang from MD Anderson Cancer
Center for valuable discussions and providing the base of skull case.
This work was supported in part by NCI Grant P01 CA21239 Proton Radiation
Therapy Research and NCI Grant R01 CA103904-01A1 Multi-criteria IMRT
Optimization. The content is solely the responsibility of the authors
and does not necessarily represent the official views of the National
Cancer Institute or the National Institutes of Health.

\section*{Appendix: Projection to satisfy the ramp constraint}

In our previous work in \cite{chen10} we have presented most of the
optimization method used in this paper. However, the ramp function
was added to the pool of objective/constraint functions. The ramp
function can be formulated as a linear problem with the help of auxiliary
variables. By introducing auxiliary variables the same projection
methods applied in \cite{chen10} can be used to handle a ramp constraint.
In the current implementation though we avoid the introduction of
auxiliary variables and use an iterative heuristic to project onto
a violated ramp constraint.

For example, we consider an underdose ramp constraint for the target.
Let $D$ be the dose-fluence matrix for the target voxels and let
the beamlet intensity vector be $x$. Then the dose vector to the
target voxels is $d=Dx$. Assume we are at a solution $x_{0}$ where
the constraint $r(Dx_{0},d^{pres})\le b$ is violated. The projection
operator needs to return a solution $P(x_{0})$ that is closest to
$x_{0}$ and satisfies the constraint. When the ramp constraint is
violated, it means some of the voxels are underdosed. Consider only
this set of underdosed target voxels. Summing up the rows of the corresponding
sub-matrix of $D$ gives the gradient of the mean dose to these voxels,
i.e., the direction of maximal increase of the mean dose of the underdosed
voxels. A move in this direction will increase the dose to these underdosed
voxels and update $x$ from $x_{0}$ to $x_{1}$. Such an incremental
move is made iteratively. During each iteration the list of underdosed
voxels and the solution are updated. This procedure is repeated, until
the ramp constraint is satisfied. The overdose ramp constraint is
treated similarly. Note that ramp objectives are handled as constraints
by the projection solver \cite{chen10}.

\end{document}